\documentclass[aps,prd,twocolumn,10pt,superscriptaddress,amsmath,amssymb]{revtex4-1}

\usepackage{graphicx}
\usepackage{subfigure}
\usepackage{epstopdf}
\usepackage{hyperref}
\usepackage{slashed}
\usepackage{bm}
\usepackage{feynmp}
\usepackage{color}
\usepackage{bbold}

\newcommand{\unitvec}{\hat{\mathbf{e}}}
\newcommand{\nablab}{\boldsymbol{\nabla}}
\newcommand{\sigmab}{\boldsymbol{\sigma}}
\newcommand{\magn}{\mathbf{m}}
\newcommand{\mmf}{\overline{m}}
\newcommand{\mfl}{\tilde{\mathbf{m}}}
\newcommand{\stag}{\mathbf{l}}

\begin{document}

\title{Topological staggered field-electric effect with bipartite magnets}

\author{Stefan Rex}
\affiliation{Department of Physics, Norwegian University of Science and Technology, N-7491 Trondheim, Norway}
\author{Flavio S. Nogueira}
\affiliation{Institute for Theoretical Solid State Physics, IFW Dresden, PF 270116, 01171 Dresden, Germany}
\affiliation{Institut f{\"u}r Theoretische Physik III, Ruhr-Universit{\"a}t Bochum, Universit{\"a}tsstra{\ss}e 150, DE-44801 Bochum, Germany}
\author{Asle Sudb{\o}}
\affiliation{Department of Physics, Norwegian University of Science and Technology, N-7491 Trondheim, Norway}

\begin{abstract}
We study the interface physics of bipartite magnetic materials deposited on a topological insulator. This comprises antiferromagnets as well as ferrimagnets and ferromagnets with multiple magnetic moments per unit cell. If an energy gap is induced in the Dirac states on the topological surface, a topological magnetoelectric effect has been predicted. Here, we show that this effect can act in opposite directions on the two components of the magnet in a certain parameter region. Consequently, an electric field will mainly generate a staggered field rather than a net magnetization in the plane. This is  relevant for the current attempts to detect the magnetoelectric effect experimentally, as well as for possible applications.
We take a field-theoretic approach that includes the quantum fluctuations of both the Dirac fermions on the topological surface as well as the fermions in the surface layer of the magnet in an analytically solvable model. 
The effective Lagrangian and the Landau-Lifshitz equation describing the interfacial magnetization dynamics are derived.
\end{abstract}

\maketitle

\section{Introduction}
Since their discovery, topological insulators (TIs) \cite{HasanKane_RMP, QiZhang_RMP} have attracted much attention due to their unique surface properties. In three-dimensional TIs, every surface exhibits linearly dispersing conducting states inside the bulk band gap. These can be described as Dirac fermions and exhibit spin-momentum locking. If time-reversal symmetry (TRS) at the surface is broken by an orthogonal net magnetization, the Dirac states become massive, i.e., a gap opens in their energy dispersion. It has been shown that this generates a Chern-Simons (CS) term in the effective field-theory which imposes a topological magnetoelectric (TME) effect \cite{QiHughesZhang_PRB,EssinMooreVanderbilt} on the surface, where an electric field induces a net in-plane magnetization. This distinct response to an electromagnetic field is a hallmark of the TI phase.

Magnetic order on the TI surface can be established by doping with $3d$ transition metals \cite{Hor2010, Chen659, KouNanoLett, KouACSNano, OkadaPRL, Chang2013, LiPRL2015a}, proximity coupling to a magnetic insulator in bilayer structures \cite{WeiPRL2013, KandalaAPL2013, YangPRB2013, MooderaNature2016, Bi2Se3-YIG}, or a combination of both \cite{LiPRL2015b}. In \cite{MooderaNature2016}, a magnetization orthogonal to the surface was realized even at room-temperature in EuS-Bi$_2$Se$_3$ bilayers. In theoretical  works, a broad range of potential applications of such heterojunctions combining ferromagnetic insulators (FMIs) and TIs have been suggested, e.g., related to spintronics \cite{GarateFranz, NomuraNagaosa, YokoyamaZangNagaosa, TserkovnyakLoss, FerreirosCortijo, LinderPRB2014, FerreirosBuijinstersKatsnelson, SemenovPRB2012, SemenovPRB2014, DuanPRB2015, FM_trilayer}, 
and several further implications of the TME effect have been discussed, including the formation of magnetic monopoles \cite{QiScience2009}, and the interplay with long-range Coulomb interaction \cite{Flavio_PRL, FM_bilayer, FM_trilayer}.

So far, not much focus has been directed at more general classes of magnetic materials. Mostly, it is assumed that the TME effect will occur in the same way as long as a net magnetization is present. However, several technologically relevant materials do not have a simple ferromagnetic (FM) structure, but are instead ferrimagnets (FiMs) or antiferromagnets (AFMs). For instance, one of the most prominent materials for spintronics devices is yttrium iron garnet (YIG), a 
FiM with a complicated crystal structure  \cite{YIG_Cherepanov, YIG_Bauer}. In YIG, an enhancement of 
the magnetization has been recently observed in a bilayer structure YIG-Bi$_2$Se$_3$, where  Bi$_2$Se$_3$ is doped with Cr \cite{Bi2Se3-YIG}. 
It is thus natural to ask if and how the topological effects will manifest itself in multicomponent FMIs, FiMs or AFM 
insulators. In AFMs, there is no net magnetization (except in some cases for special surface orientations \cite{LuoPRB2013}). However, 
a gap can still be opened at the Dirac points, as in the FM and FiM cases, by means of magnetic doping in the TI. 
Such a system has recently been realized experimentally \cite{HKG16}.

In the present paper, we study a bilayer heterostructure consisting of a bipartite magnetic insulator (BMI) and a TI. We show that, depending on the microscopic parameters of the BMI, the TME effect can take the opposite sign on the two sublattices, turning the overall electric-field response from a TME effect into a topological staggered-field (TSE) effect. Our calculation is to be understood as a proof of principle, as the model we employ is  
simplified and may not suffice to make quantitative predictions. On the other hand, we are able to obtain fully analytic solutions within a field-theoretic approach that accounts for the fermionic quantum fluctuations on both the BMI and the TI surfaces. We will derive the effective Lagrangian, revealing the structure of the magnetoelectric response, and the Landau-Lifshitz equation (LLE) of the interfacial magnetization dynamics. We work in Gaussian units and set $\hbar=1$. All calculations are done at zero-temperature. This is justified as long as the Fermi level is tuned to lie in the induced energy gap, for instance by gating of the interface.

The model we use is described in the following section. We discuss the non-topological fluctuation effects originating with the electrons on the BMI surface in section~\ref{Sec:IntegrateChi}, before we move on to the topological effects that are revealed upon integrating out the Dirac states in section~\ref{Sec:TSE}. We summarize our results in section~\ref{Sec:Conclusion}.

\section{Model system}\label{SecModel}
\begin{figure}
\includegraphics[width=\columnwidth]{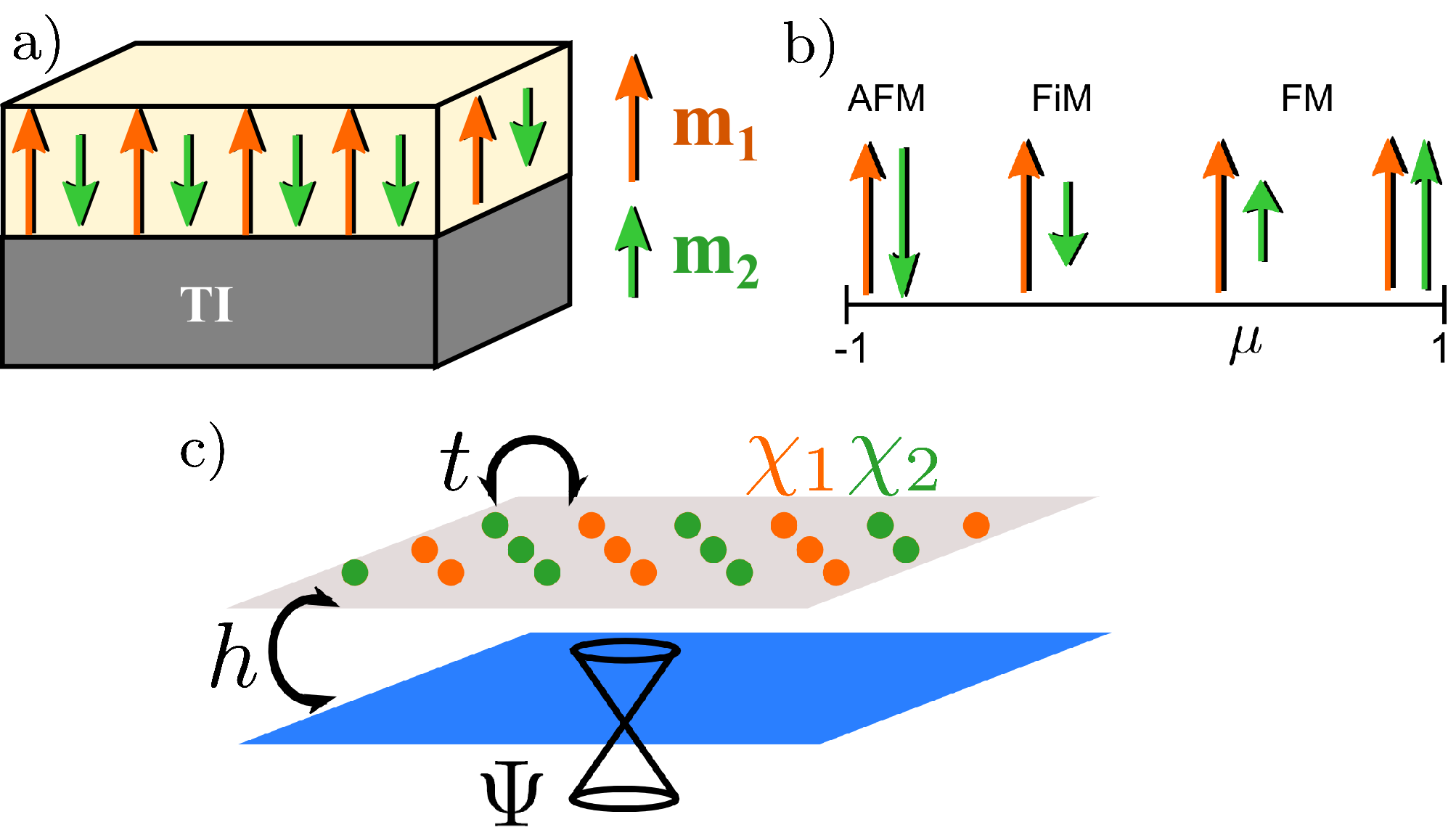}
\caption{(Color online) The model system: a) Bilayer heterostructure consisting of a bipartite magnetic insulator (BMI) deposited on a topological insulator (TI). b) By means of the parameter $\mu=\overline{m}_2/\overline{m}_1$, the magnet can be tuned to be in an antiferromagnetic (AFM), ferrimagnetic (FiM), or ferromagnetic (FM) configuration at mean-field. c) The model involves fermionic fields $\Psi$ and $\chi_{1,2}$ on the surfaces of both the TI (blue plane) and the BMI (grey plane), respectively, which are coupled by the amplitudes $h$ (hopping across the interface) and $t$ (local coupling of the two sublattices).}
\label{Fig:Model}
\end{figure}

Describing the heterostructure, one has to account for the contributions from the bulk of the BMI, the surfaces of the BMI and the TI, hopping across the 
interface due to proximity, and Coulomb interactions between the Dirac electrons at the interface. The bulk of the TI is required to guarantee the 
existence of the topological surface states, but does not appear explicitly.
The model system is illustrated in Fig.~\ref{Fig:Model}.

We start with the surface of the TI which is chosen to be the $(x,y)$-plane and described by the Dirac Lagrangian
\begin{equation}
\mathcal{L}_\text{D} = \Psi^\dagger[i\partial_t-iv_F(\sigma_y\partial_x-\sigma_x\partial_y) + e(\varphi+\phi)]\Psi
\label{Dirac_1}
,\end{equation}
where $\Psi=[\psi_\uparrow~ \psi_\downarrow]^T$ are the surface Dirac fermions, $v_F$ is the Fermi velocity, $\varphi$ the fluctuating potential of Coulomb interactions among the Dirac fermions, and $\phi$ is any externally applied electric potential. 
A term quadratic in $\varphi$ describes the Coulomb interaction in the plane \cite{Flavio_PRL, FM_bilayer, FM_trilayer}:
\begin{equation}
\mathcal{L}_\text{Cou}(\mathbf{r}) = -\frac{1}{8\pi^2}[\nablab_\parallel\varphi(\mathbf{r})]\cdot\int d^2r^\prime\frac{\nablab_\parallel^{\prime}\varphi(\mathbf{r}^\prime)}{|\mathbf{r}-\mathbf{r}^\prime|},
\end{equation}
where $\nablab_\parallel=(\partial_x,\partial_y)$ denotes the in-plane gradient operator. 

We model the bulk bipartite magnetic material as two interpenetrating FMs (denoted by indices $i=1,2$) that are coupled by an exchange interaction,
$\mathcal{L}_\text{bulk} = \mathcal{L}_1 + \mathcal{L}_2 + \mathcal{L}_\text{ex}$, where
\begin{equation}
\label{Eq:L-Mag}
\mathcal{L}_i = -\textbf{b}(\magn_i)\cdot\partial_t\magn_i - \frac{\kappa}{2}(\nablab\magn_i)^2
\end{equation}
and
\begin{equation}
\mathcal{L}_\text{ex}(\mathbf{r}) = -\lambda\magn_1(\mathbf{r})\cdot\magn_2(\mathbf{r}).\label{Eq:Lexchange}
\end{equation}
Here, $\mathbf{b}$ is the Berry connection, which satisfies $\nablab_{\magn_i}\times\mathbf{b}(\magn_i) = \magn_i/m_i^2$, $\kappa>0$ is the FM exchange energy, and $\lambda>0$ ($<0$) for AFM (FM) coupling of the two components. In the bulk model, we ignore anisotropy terms. It will turn out that the system intrinsically contains anisotropy, and additional bulk contributions would not qualitatively alter the physics.

In order to describe the surface Berry phases associated to the two sublattices, we introduce fermionic fields 
$\chi_i=[\chi_{i\uparrow}~\chi_{i\downarrow}]^T$, $i=1,2$ representing sublattice indices, which when integrated out generate the desired 
surface Berry phases. This procedure to generate Berry phases is well known in the literature \cite{Stone-WZW-term,Fujikawa-2005}, 
and is very useful in our case because it permits coupling the underlying sublattice fermions to the Dirac surface states.  
The surface layer of the bipartite magnetic insulator is thus described by the Hamiltonian,
\begin{eqnarray}
{\cal H}_{\rm surf}=-t(\chi_1^\dagger\chi_2+\chi_2^\dagger\chi_1)-J\sum_{i=1,2}{\bf m}_i\cdot\chi_i^\dagger\sigmab\chi_i
\label{Chi_Mag_Surface}
,\end{eqnarray}
where $J$ the strength of the exchange coupling to the respective magnetization $\magn_i(z=0)$, $\sigmab$ are the Pauli matrices, and $t$ a 
paramater coupling the surface fermions of the BMI on different sublattices.  It will be crucial in obtaining a TSE, and
also leads to mixed Berry phase terms originating on the different sublattices.  When $t=0$, the surface Berry phases decouple and 
just correspond to a shift of the Berry phases already present in Eq. (\ref{Eq:L-Mag}). 
Note that the Lagrangian only accounts for coupling of fermions $\chi_1$  and $\chi_2$  within one unit cell, thus being momentum-independent 
in the continuum limit. Further electron dynamics (gradient terms) is neglected. This rough approximation is valid as long as the magnet is a 
strong insulator and the gap is much larger than the induced gap in the Dirac states. It does not spoil the generation of the surface Berry 
phases, however.  Furthermore, the lattice model of the surface of the magnet does not explicitly include nearest-neighbor exchange interactions, 
which are already captured by the Lagrangian of the magnetic bulk. The chemical potential is set to zero for the electrons on both surfaces 
because the Fermi level is assumed to be tuned to lie in the gap.

If the surfaces of the TI and the AFM or FiM are in proximity to each other, there is also an amplitude $h$ that couples the 
surface fermions of the magnetic insulator to the surface fermions of the topological insulator, 
\begin{equation}
\mathcal{L}_\text{int} = h[\Psi^\dagger(\chi_1+\chi_2)+(\chi_1^\dagger+\chi_2^\dagger)\Psi].
\end{equation}
Our calculation amounts to integrating out all fermionic fields in order to obtain an effective theory of the magnetization.

\section{Quantum fluctuations of the sublattice fermions}\label{Sec:IntegrateChi}
We start by integrating out the fermions $\chi_i$ of the BMI surface to obtain an effective model for the Dirac fermions $\Psi$. We assume that the mean-field direction of the magnetization is orthogonal to the interface, such that a mass in the Dirac states can be induced. 
We write $\magn_i^\text{mf}=\mmf_i\unitvec_z$ and define the dimensionless parameter $\mu=\mmf_2/\mmf_1$, where without loss of generality 
$|\mu|\leq 1$. Then, $\mu>0$ describes a FM, $-1<\mu<0$ a FiM, and $\mu=-1$ an AFM (Fig.~\ref{Fig:Model}b). We also introduce $\tau=t^2/J^2\mmf_1^2$, which will be useful later.
From Eq. \ref{Chi_Mag_Surface}, we define a matrix
\begin{equation}
A = \begin{pmatrix}i\partial_t+ J\magn_1\cdot\sigmab&t\\t&i\partial_t+ J\magn_2\cdot\sigmab\end{pmatrix}
\end{equation}
such that the action of the surface of the magnetic insulator is symbolically written as $\mathcal{S}_\text{surf} = \chi^\dagger A  \chi$, 
where $\chi^\dagger=(\chi_1^\dagger,\chi_2^\dagger)$. The integral over spacetime is implicit in this symbolic representation. We use a spinor 
$\tilde \Psi^\dagger=(\Psi^\dagger,\Psi^\dagger)$ that contains the same Dirac fermion twice, to write 
$\mathcal{L}_\text{int}=h\chi^\dagger \tilde \Psi + \text{h.c.}$ We next proceed by integrating out the magnetic 
surface fermions $\chi$, 
\begin{eqnarray}
\mathcal{Z} & = & \int\! D\,[\overline{\chi},\chi]\,e^{i\int dt\int d^2r(\mathcal{L}_\text{surf}+\mathcal{L}_\text{int})} \nonumber \\
& = &\int\! D\,[\overline{\chi},\chi]\,e^{i(\chi^\dagger A \chi-h \chi^\dagger \tilde \Psi -h \tilde \Psi^\dagger \chi)} \nonumber \\
&= & \exp\left(i\text{Tr}\,\ln A+ih^2 \tilde \Psi^\dagger A^{-1} \tilde \Psi\right).
\label{Eq:GaussianIntegral}
\end{eqnarray}
Note that the notation $\text{Tr}$ also contains the integration over 
the quantum numbers besides the matrix trace. We will discuss the two terms in the last line separately in the following subsections.

\subsection{Surface corrections to the bulk terms}\label{Sec:NonTop} 
The term $\text{Tr}\ln\,A$ in Eq.~\eqref{Eq:GaussianIntegral} is independent of the topological Dirac states. It leads to the Berry phases mentioned previously and renormalizes the magnetic bulk terms at the surface. Details of the calculation and complete analytical expressions can be found in 
Appendix~\ref{App:Anisotropy}. We finally obtain
\begin{eqnarray}
&&\delta\mathcal{L}_\text{mag}(\mathbf{r},t) =\notag\\
&&-2J^2\magn_1\cdot\text{diag}(T^{00}-T^{zz},T^{00}-T^{zz},T^{00}+T^{zz})\cdot\magn_2\notag\\
&&+2J^2\sum_{i=1,2}\Big\{\left[(D_i^{00}+D_i^{zz})\mmf_i + (T^{00}+T^{zz})\mmf_{3-i}\right]m_{iz}\notag\\
&&{}- D_i^{zz}m_{iz}^2 +\mathcal{D}^{0z}_i\unitvec_z\cdot\left[\magn_i(\mathbf{r},t)\times\partial_t\magn_i(\mathbf{r},t)\right]\Big\}\notag\\
&&{}+2J^2\mathcal{T}^{0z}\unitvec_z\cdot\big[\magn_1(\mathbf{r},t)\times\partial_t\magn_2(\mathbf{r},t)\notag\\
&&{}+\magn_2(\mathbf{r},t)\times\partial_t\magn_1(\mathbf{r},t)\big]
\label{Eq:MagLagr}
,\end{eqnarray}
where $D_i^{00}, D_i^{zz}, \mathcal{D}^{0z}_i, T^{00}, \mathcal{T}^{0z}$ and $T^{zz}$ are functions of $t,J,\mmf_i$ and the lattice spacing $a$. The Berry phases are represented by the cross-product terms. The terms proportional to $\mathcal{D}^{0z}_i$ shift the Berry phases introduced in Eq.~\eqref{Eq:L-Mag}, while the term proportional to $\mathcal{T}^{0z}$ is a mixed Berry phase term. We remark that $\mathcal{T}^{0z}\propto t$, thus no mixed Berry phase appears if $t=0$.

\begin{figure}
\includegraphics[width=\columnwidth]{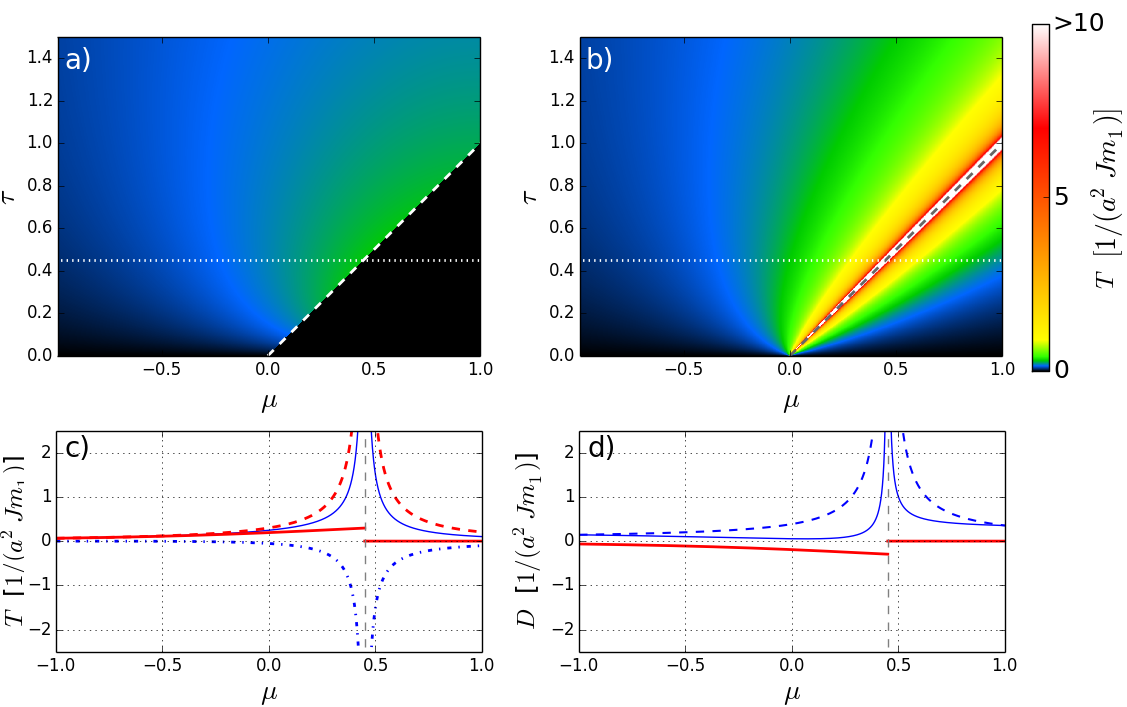}
\caption{(Color online) The anisotropic fluctuation-induced antiferromagnetic exchange coupling of $\magn_1$ and $\magn_2$ at the surface, which renormalizes the exchange coupling induced from the bulk. a) In the component along the mean-field direction, the coupling constant is given by $T^{00}+T^{zz}$ (see main text) and shows a finite discontinuity at $\mu=\tau$ (dashed line). b) In the component orthogonal to the mean-field direction, the AFM coupling $T^{00}-T^{zz}$ diverges at the discontinuity. The color scale is identical in both plots. c) The quantities $T^{00}$ (thin solid blue line), $T^{zz}$ (dash-dotted blue line), $T^{00}+T^{zz}$ (bold solid red line), and $T^{00}-T^{zz}$ (dashed red line) as a function of $\mu$ for a specific value of $\tau$ ($\tau=0.45$), which is indicated by the thin white dotted lines in plots a) and b). d) The anisotropy terms $D_1^{zz}$ (thin solid blue line), $D_2^{zz}$ (dashed blue line), and $D_i^{00}+D_i^{zz}$ (bold solid red line, identical for $i=1,2$) behave similarly, showing a discontinuity at $\tau=\mu$. The vicinity of this line is excluded from the further analysis.}
\label{Fig:Discontinuity}
\end{figure}

Furthermore, the coupling of $\magn_1$ and $\magn_2$ given by Eq.~\eqref{Eq:Lexchange} is renormalized by the first line in Eq.~\eqref{Eq:MagLagr} and becomes anisotropic. This leads to in-plane and out-of-plane effective exchange couplings given by, 
\begin{equation}
\lambda_{\rm eff}^\parallel=\lambda+2J^2(T^{00}-T^{zz}),
\end{equation}
\begin{equation}
\lambda_{\rm eff}^\perp=\lambda+2J^2(T^{00}+T^{zz}).
\end{equation}
An evaluation of our analytic expressions (Appendix ~\ref{App:Anisotropy}) reveals that the dynamically generated coupling favors AFM alignment of the two magnetic components. Indeed, using Eqs. (\ref{Eq:T00}) and (\ref{Eq:Tzz}) of Appendix A, we obtain,
\begin{equation}
\label{Eq:T00-Tzz}
T^{00}-T^{zz}=\frac{t^2[2|t^2-J^2\mmf_1\mmf_2|+2t^2+J^2(\mmf_1^2+\mmf_2^2)]}{2a^2|t^2-J^2\mmf_1\mmf_2|(M_++M_-)^3},
\end{equation}
\begin{equation}
\label{Eq:T00+Tzz}
T^{00}+T^{zz}=\frac{t^2[1+{\rm sgn}(t^2-J^2\mmf_1\mmf_2)]}{a ^2(M_++M_-)^3},
\end{equation}
where,
\begin{eqnarray}
M_\pm^2&=&\frac{J^2}{2}(\mmf_1^2+\mmf_2^2)+t^2
\nonumber\\
&\pm&\frac{J^2}{2}|\mmf_1+\mmf_2|\sqrt{(\mmf_1-\mmf_2)^2+\left(\frac{2t}{J}\right)^2}.
\end{eqnarray}
The coupling constants show a discontinuity at $t^2=J^2\mmf_1\mmf_2$, or
$\tau=\mu$,  as shown in Fig. \ref{Fig:Discontinuity}. Indeed, we see that Eq. (\ref{Eq:T00-Tzz}) diverges for $t^2=J^2\mmf_1\mmf_2$,  
	while (\ref{Eq:T00+Tzz}) vanishes when $t^2<J^2\mmf_1\mmf_2$. This divergence obviously does not occur when $\mmf_1\mmf_2<0$, 
	corresponding to the AFM case, further corroborating the favoring of the AFM alignment. Physically, the divergence for 
	$\tau=\mu$ implies the vanishing of the in-plane susceptibility.

The remaining terms in Eq.~\eqref{Eq:MagLagr} describe a z-axis anisotropy in both magnetizations. As we mentioned in section~\ref{SecModel}, our model does not account for possible anisotropy contributions originating with the bulk of the magnet. Such terms would simply be renormalized by the corresponding coefficients in Eq~\eqref{Eq:MagLagr} without changing the physical picture.

Our view of the dynamically generated surface terms as corrections to the bulk values will only hold as long as the surface effects are not too large. As can be seen from Fig.~\ref{Fig:Discontinuity}, within our model some surface terms are divergent at the discontinuity at $\mu=\tau$. Therefore, the vicinity of this line in parameter space will be excluded in our further analysis.

As a side remark, the fluctuation effects discussed in this subsection can easily be generalized to account for magnetizations that are, at mean-field, tilted relative to the surface. We have checked that Eq.~\eqref{Eq:MagLagr} remains valid when the $z$ components are replaced by mean-field components in an arbitrary direction.

\subsection{Effective Dirac Lagrangian}
The term $h^2 \tilde \Psi^\dagger A^{-1} \tilde \Psi$ in Eq.~\eqref{Eq:GaussianIntegral} may now be added to Eq.~\eqref{Dirac_1} to yield an effective action for the Dirac fermions 
\begin{equation}
\mathcal{S}_\text{eff} = \int dt\int d^2 r{\cal L}_{\rm eff}= \int dt\int d^2 r\left(\mathcal{L}_\text{D} + h^2\tilde{\Psi}^\dagger A^{-1} \tilde{\Psi}\right).
\end{equation}
Multiplying out $\tilde \Psi^\dagger A^{-1}\tilde \Psi$ into single-fermion operators 
again, we find the effective Lagrangian of the Dirac electrons at the coupled surfaces,
\begin{eqnarray}
\mathcal{L}_\text{eff} &=& \mathcal{L}_\text{D} + \gamma\Psi^\dagger\left(\frac{t^2}{J^2}-\magn_1\cdot\magn_2\right)\Psi \notag\\
&&{}+\Psi^\dagger\left(J_1\magn_1\cdot\sigmab+J_2\magn_2\cdot\sigmab\right)\Psi,
\label{Eq:EffDiracLagrangian1}
\end{eqnarray}
where we have defined the constant
\begin{equation}
\gamma = \frac{2th^2J^2}{\text{det}\,A}
\end{equation}
and the effective magnetic coupling constants for the two sublattices
\begin{equation}\label{Eq:Jconstant}
J_i = \frac{h^2J}{\text{det}\,A}\left(J^2\magn_{3-i}^2-t^2\right),
\end{equation}
where 
\begin{eqnarray}
\text{det}\,A & = & (-\partial^2_t - t^2)^2 + J^2 \partial^2_t (\magn_1^2 + \magn_2^2) \nonumber \\
& + & J^2( J^2 \magn_1^2 \magn_2^2 - 2 t^2 \magn_1\cdot\magn_2) . 
\end{eqnarray} 
In $\text{det}\,A$, the fluctuations in $\magn_{1,2}$ are not of leading order. Therefore, we will approximate the determinant in the Dirac Lagrangian by 
its mean-field value $\text{det}\,A^\text{mf} =  t^4 + J^2 \left[J^2 \mmf_1^2 \mmf_2^2 - 2 t^2 \mmf_1 \mmf_2 \right]$, whereby we also 
neglected higher-order time derivatives in the low-frequency limit. Furthermore, we assume that the coupling $h$ of the
surface fermions $\chi$ and $\Psi$ at the interface is small compared to the internal energy scales of the 
magnet, $t$ and $J\mmf_i$. 
Otherwise, one obtains a renormalization of the time scale. It is interesting to note that the term $\propto\gamma$ in 
Eq.~\eqref{Eq:EffDiracLagrangian1} contributes to the chemical potential of $\Psi$. The chemical potential may be tuned by 
means by adjusting $\phi$ appearing in Eq. \eqref{Dirac_1}, and the mean-field part of the second term in Eq. \eqref{Eq:EffDiracLagrangian1} 
may thus always be adjusted away. We will only keep the remainder to linear order in the fluctuations.

Note that the sign of $J_i$ in Eq.~\eqref{Eq:Jconstant} depends on the parameter $t$ appearing in Eq. \eqref{Chi_Mag_Surface}, as well as the magnitude of the magnetic moments. This is a key observation that we will return to when discussing the topological effects in the next section.

\section{Topological magnetoelectric effects}\label{Sec:TSE}
Now, we express the effective Lagrangian Eq.~\eqref{Eq:EffDiracLagrangian1} as
\begin{equation}
\mathcal{L}_\text{eff} = \overline{\Psi} (i\slashed\partial + m_\Psi)\Psi + \overline{\Psi} (\tilde{\sigma}-\slashed{a})\Psi
\label{Eq:EffDiracLagrangian2}
,\end{equation}
where the first term is the mean-field part with $\partial=(\partial_t,v_F\nabla_\parallel)$ and $m_\Psi=J_1\mmf_1+J_2\mmf_2$, whereas the second term contains the fluctuating fields $\tilde{\sigma}=J_1\tilde{m}_{1z}+J_2\tilde{m_2}_z$ and
\begin{equation}
\mathbf{a} = \begin{pmatrix}-e(\varphi+\phi)+\gamma(\mmf_1\tilde{m}_{2z}+\mmf_2\tilde{m}_{1z})\\J_1\tilde{m}_{1y}+J_2\tilde{m}_{2y}\\-J_1\tilde{m}_{1x}-J_2\tilde{m}_{2x}\end{pmatrix}
.\end{equation}
From this representation, one can see that the out-of-plane fluctuations of the magnetization contribute to the effective electric potential at the interface. This is a result of the fluctuations in the chemical potential that we have observed in Eq.~\eqref{Eq:EffDiracLagrangian1}.
To obtain an effective field theory for the magnetizations that contains the proximity effects induced by the topological insulator, we also have to integrate out the remaining fermions $\Psi$ and the fluctuating Coulomb potential $\varphi$. Equation~\eqref{Eq:EffDiracLagrangian2} is formally equivalent to the field theory studied in Refs.~\onlinecite{Flavio_PRL, FM_bilayer}, given that the mass term $m_\Psi$ is nonzero. This is naturally the case for FMs and FiMs (except at $\mu=\tau$, which we already excluded), while it might be enforced by doping in the case of an AFM.

Integrating out $\Psi$ yields the fluctuation-induced La\-gran\-gian to one-loop order in the vacuum polarization diagrams \cite{Flavio_PRL, FM_bilayer},
\begin{equation}
\delta\mathcal{L}_\text{eff} = 
\frac{\epsilon_{\mu\nu\lambda}a^\mu\partial^\nu a^\lambda}{8\pi}
-\frac{(\epsilon_{\mu\nu\lambda}\partial^\nu a^\lambda)^2}{24\pi m_\Psi}
-\frac{m_\Psi\tilde{\sigma}^2}{2\pi}
+\frac{(\partial\tilde{\sigma})^2}{24\pi m_\Psi}
\label{EqLagrangianNoFermion}
\end{equation}
The first term is the CS term that is responsible for all topologically protected contributions to the Lagrangian. The other terms correspond to a Maxwell term and out-of-plane anisotropy.

Besides these dynamical terms, a term describing the energy at mean-field is produced after all fermionic fields have been integrated out. This term can be expanded into a Landau theory for the mean-field magnetizations at the BMI-TI interface. The Landau expansion can be found in App.~\ref{App:LandauExpansion}, where we find that the quadratic term is always negative. This serves as a check that our model, where we treated $\mmf_{1,2}$ as parameters, is consistent with the existence of a magnetic phase.

Reinserting $\mathbf{a}$, we can separate $\delta\mathcal{L}_\text{eff}$ in a Coulomb-interaction ($\varphi$-dependent) part $\mathcal{L}_\varphi$ and the remaining dynamically generated terms $\mathcal{L}_\text{dyn}$.
After integrating out $\varphi$, the Coulomb contributions become
\begin{equation}
\mathcal{L}_\varphi(\mathbf{r},t) = 2\rho(\mathbf{r},t)\int\! d^2r^\prime\frac{\rho(\mathbf{r}^\prime,t)}{|\mathbf{r}-\mathbf{r}^\prime|}
,\end{equation}
with the charge density
\begin{eqnarray}
\rho &=& \frac{e}{8\pi v_F}\nablab_\parallel\cdot(J_1\magn_1+J_2\magn_2)+\frac{e^2}{24\pi m_\Psi}\nablab_\parallel \mathbf{E}_\text{ext}\notag\\
&&{}-\frac{e}{24\pi m_\Psi v_F}\left[\nablab_\parallel \times\partial_t(J_1\magn_1+J_2\magn_2)\right]\cdot\unitvec_z\notag\\
&&{}+\frac{\gamma e}{24\pi m_\Psi}\left(\nablab_\parallel \right)^2(\mmf_1 m_{2z}+\mmf_2 m_{1z}),
\label{Eq:ChargeDensity}
\end{eqnarray}
where $\mathbf{E}_\text{ext}=-\nablab\phi$ is the externally applied electric field. We also define the Coulomb field induced by the charge density,
\begin{equation}
\mathbf{E}_\text{Cou}(\mathbf{r}) = -\int\!d^2r^\prime\,\frac{\mathbf{r}-\mathbf{r}^\prime}{|\mathbf{r}-\mathbf{r}^\prime|^3}\rho(\mathbf{r}^\prime).
\end{equation}
For low frequency and momentum, the last two terms in Eq.~\eqref{Eq:ChargeDensity} will be negligible compared to the first two terms.

The part of the Lagrangian that is due to the non-trivial topology (i.e., stemming from the CS term), where we write $\mathbf{M}=J_1\magn_1+J_2\magn_2$ for brevity, can be expressed explicitly as
\begin{eqnarray}
\mathcal{L}_\text{topol} &=&
\frac{e}{4\pi v_F}\mathbf{M}_\shortparallel\cdot(\mathbf{E}_\text{ext}
+\mathbf{E}_\text{Cou})\notag\\
&&{}- \frac{1}{8\pi v_F^2}\left(\mathbf{M}\times\partial_t\mathbf{M}\right)\cdot\unitvec_z\notag\\
&&{}+\frac{\gamma}{4\pi v_F}\mathbf{M}\cdot\nablab_\parallel\left(\mmf_1 m_{2z}+\mmf_2 m_{1z}\right).
\label{Eq:LagTopol}
\end{eqnarray}
The first term represents the magnetoelectric coupling, involving both the external field and the fluctuation-induced Coulomb field. The second term is a Berry phase. Unlike the Berry phase generated by the fluctuations of $\chi$, this expression always includes mixed terms, regardless of the parameter $t$. Finally we also obtain a topological coupling of the magnetic in-plane and out-of-plane fluctuations.

\begin{figure}
\includegraphics[width=0.4\columnwidth]{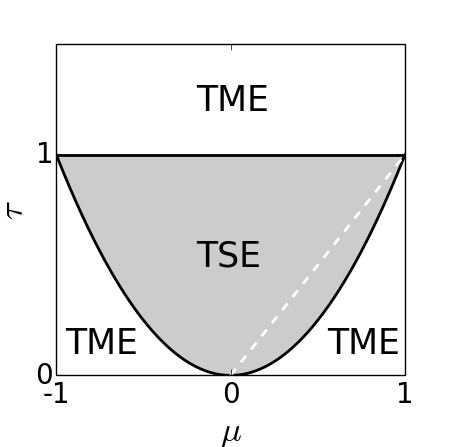}
\includegraphics[width=0.58\columnwidth]{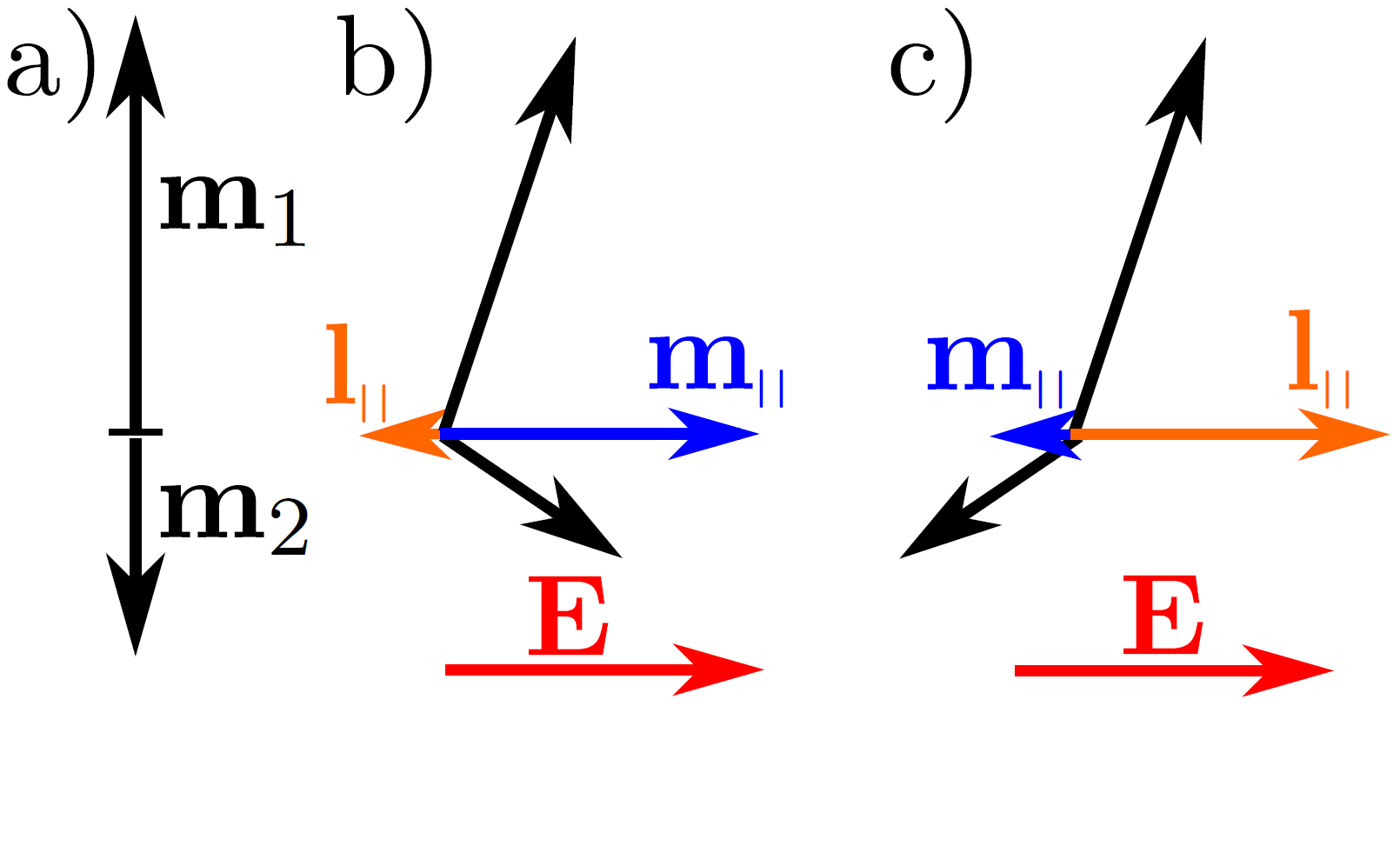}
\caption{(Color online) Left: parameter regions of the bipartite magnet where the topological response to an electric field has the same (white area) or opposite (grey area) direction on the two sublattices, corresponding to a predominantly magneto-electric (TME) or staggered field-electric (TSE) effect, respectively. Here, $\tau$ is the  dimensionless amplitude of the coupling of the fermions on the two sublattices and $\mu$ is the ratio of the mean-field values of the magnetizations on the sublattices. Close to the dashed line at $\mu=\tau$, our results may not be applicable. Right: illustration of the topological effects for a FiM: a) both magnetizations $\magn_1$ and $\magn_2$ (black) pointing in their mean-field directions. b) TME effect: if the topological response to the electric field $\mathbf{E}$ (red) has the same sign on both sublattices, an in-plane net magnetization $\magn_\shortparallel$ (blue) is generated, while the induced in-plane staggered field $\stag_\parallel$ (orange) is small. c) TSE effect: if the topological response to $\mathbf{e}$ has opposite signs for $\magn_1$ and $\magn_2$, an in-plane staggered field is generated, while $\magn_\shortparallel$ is small. The overall sign of these effects depends on the sign of the mass term $m_\Psi$.}
\label{Fig:TME}
\end{figure}

At this point, we can discuss how the system will respond to an electric field. This is the main result of our paper. As we can see from Eq.~\eqref{Eq:LagTopol}, the electric field is coupled to $\mathbf{M}$ in the same way as it couples to the magnetic polarization in the usual TME effect. Now, let us write $\mathbf{M}$ in terms of the net magnetization $\magn=\magn_1+\magn_2$ and the staggered field $\stag=\magn_1-\magn_2$,
\begin{equation}
\mathbf{M} = \frac12(J_1+J_2)\magn + \frac12(J_1-J_2)\stag
.\end{equation}
Obviously, if $J_1$ and $J_2$ have the same sign, an electric field will mainly generate a net in-plane magnetization, while the coupling to the staggered field is small. Overall, the system will behave as one would expect for a simple FM. However, if $J_1$ and $J_2$ have opposite signs, an electric field will mainly induce a staggered field in the plane, while the response in the net magnitization will be weak. This is because the usual TME effect takes place on both sublattices, but with opposite direction. Going back to Eq.~\eqref{Eq:Jconstant}, it is easy to find the parameter region where this topological staggered field-electric (TSE) effect can be found. In terms of the dimensionless model parameters, the condition for $J_1$ and $J_2$ having opposite signs is $\mu^2<\tau<1$, see figure~\ref{Fig:TME}. A purely TSE response is expected if $J_1=-J_2$, which is the case if $\tau=\frac{1}{2}(1+\mu^2)$. Remarkably, the predominantly TSE response can appear even in a FM material ($\mu>0$), if it consists of multiple magnetic components per unit cell with different magnitude and a suitable parameter $t$. Thus, it is possible that experiments fail to detect the usual TME effect even when a decent gap opening occurs. In contrast, a purely AFM material ($\mu=-1$) would not show any coupling to the staggered field, even in the presence of a mass term $m_\Psi$ by magnetic doping, because $J_1=J_2$ for equally strong magnetic moments on the two sublattices. Our model of the BMI is quite simple, and for a real material it might be much harder to find the parameter regions that allow for the observation of the TME or TSE effect. However, it is a remarkable finding that the overall topological response in a BMI-TI heterostructure can depend dramatically on microscopic details of the magnet.

A restriction on our findings is imposed by the discontinuity discussed in the previous section. Due to divergent terms, our results on the TSE effect will not be applicable for parameters in the vicinity of the line $\mu=\tau$ in Fig.~\ref{Fig:TME}.

Previous work has found a Coulomb-mediated magnetic dipolar interaction \cite{FM_trilayer}. The Coulomb interaction in the present work will lead to the same effect within each sublattice. Moreover, there will be a dipolar interaction between the sublattices. Again, for a system in the TSE regime, we will get an effect in the opposite direction. Thus, the inter-component dipolar interaction will favour counteralignment instead of alignment of $\magn_{1,\shortparallel}$ and $\magn_{2,\shortparallel}$.

Our model also reveals a topological coupling of the in-plane components of the magnetic moments and the gradient in the out-of-plane component as described by the last term in Eq.~\eqref{Eq:LagTopol}, which can be understood as an anomalous spin-stiffness term. This term has not been considered in previous studies and can lead to a spin canting effect if the magnetization is not homogeneous, as in the presence of spin-waves or domain walls. For the observation of the electromagnetic response it will, however, not be important.

The full Lagrangian describing the magnetic moments in the system is now given by
\begin{equation}
\mathcal{L}_\text{tot} = \mathcal{L}_\text{bulk} + \mathcal{L}_\varphi + \mathcal{L}_\text{dyn} + \delta\mathcal{L}_\text{mag},\label{Eq:Lmag_tot}
\end{equation}
from which the coupled LLEs for the motion of $\magn_1$ and $\magn_2$ at the interface can be derived. It takes the form
\begin{equation}
\partial_t \begin{pmatrix} \magn_1 \\ \magn_2 \end{pmatrix} =
\Xi^{-1} \begin{pmatrix} \magn_1\times\mathbf{d}_1 \\ \magn_2\times\mathbf{d}_2 \end{pmatrix}
.\end{equation}
For details, we refer to Appendix~\ref{App:LLE}. The $(6\times6)$ matrix $\Xi$ contains all Berry phase terms. In particular, there are off-diagonal terms that stem from the fluctuation-induced mixed Berry phases. Such terms are generated by the fermions $\chi_i$ (if $t\neq0$) as well as the Dirac fermions $\Psi$. The contribution by the fluctuations of $\Psi$ is of topological origin, as it stems from the CS term. 
The effective fields $\mathbf{d}_i$ contain, besides spin-stiffness and anisotropy terms, a topological part
\begin{eqnarray}
\mathbf{d}^i_\text{topol} &=&
\frac{eJ_i}{4\pi v_F}\mathbf{E}_\text{Cou}
+ \frac{eJ_i}{4\pi v_F}\mathbf{E}_\text{ext}
-\frac{\gamma\mmf_{3-i}}{4\pi v_F}(\nabla_\parallel\cdot\mathbf{M})\unitvec_z\notag\\
&&{}-\frac{\gamma J_i}{4\pi v_F}\nabla_\parallel(\mmf_1 m_{2z} + \mmf_2 m_{1z})
\end{eqnarray}
corresponding to Eq.~\eqref{Eq:LagTopol}. The first two terms show explicitly how the external electric field and the Coulomb field affect the magnetization dynamics as a consequence of the magnetoelectric effects discussed above.

\section{Conclusion}\label{Sec:Conclusion}
We have studied the topological effects at the interface of a TI and a BMI within an analytically accessible model that accounts for the fermionic quantum fluctuations at the surfaces of both materials. We have demonstrated that the TME effect that is known for magnetic TI surfaces can take the opposite 
sign for the different magnetic components, depending on microscopic details of the material. This leads to an overall TSE response to an electric field, while 
the induced net magnetization in the plane can be weak even in the presence of a stable energy gap in the Dirac dispersion. Thus, experiments that aim at detecting 
the TME effect might also look for a response in the staggered field. A response in the magnetization can be absent even when a FM insulator is used, if there are multiple magnetic components with different magnitude. 
In addition to the TSE effect, we have derived several dynamically generated Berry phases, including terms mixing $\magn_1$ and $\magn_2$. 
We also found a topological coupling of in-plane and out-of-plane magnetic components which is present  for non-homogeneous magnetization.
The fluctuations of the fermions on the magnets' surface cause discontinuity in our model, close to which our results are not applicable.

\section*{Acknowledgements}
S. R. and A. S. acknowledge funding by the Norwegian Research Council, Grants No. 205591/V20 and No. 216700/F20.

\appendix
\section{Derivation of the surface corrections}\label{App:Anisotropy}
Here, we derive the magnetic surface terms discussed in Sec.~\ref{Sec:NonTop} that are generated by $\text{Tr}\,\ln A$ in the Gaussian integral, Eq.~\eqref{Eq:GaussianIntegral}. Splitting $A=A^\text{mf}+A^\text{fl}$ in the mean-field part and the quantum fluctuations,
\begin{equation}
A^\text{mf} = \begin{pmatrix}i\partial_t+J\mmf_1\sigma_z&t\\t&i\partial_t+J\mmf_2\sigma_z\end{pmatrix}
\end{equation}
\begin{equation}
A^\text{fl} = \begin{pmatrix}J\mfl_1\cdot\sigmab&0\\0&J\mfl_2\cdot\sigmab\end{pmatrix}
,\end{equation}
we obtain the usual expansion
\begin{equation}
\text{Tr}\,\ln A = \text{Tr}\,\ln A^\text{mf} - \frac12\text{Tr}\left(GA^\text{fl}\right)^2
\label{Eq:MagTrace}
\end{equation}
where the first term is a constant corresponding to the ground-state energy that will be dealt with in App.~\ref{App:LandauExpansion}, while the second term describes the dynamics close to equilibrium to leading order. The propagator $G$ is given by $\left(A^\text{mf}\right)^{-1}$. In reciprocal space and imaginary time, $G$ depends only on the frequency $\omega$ but not on momentum, because the hopping terms in our model are momentum-independent. For all momentum integrals, we use $\pi/a$ as a cut-off value, where $a$ is the lattice spacing. The propagator can be written in the form
\begin{equation}
G = \frac{1}{\text{det}\,A^{\text{mf}}}\begin{pmatrix}D_1^0+D_1^z\sigma_z&T^0+T^z\sigma_z\\T^0+T^z\sigma_z&D_2^0+D_2^z\sigma_z\end{pmatrix}
,\end{equation}
where the components are
\begin{eqnarray}
D_1^0(\omega) &=& i\omega^3+i\omega J^2\mmf_2^2 + i\omega t^2\\
D_2^0(\omega) &=& i\omega^3+i\omega J^2\mmf_1^2 + i\omega t^2\\
D_1^z(\omega) &=& J\mmf_1\omega^2 + J^3\mmf_2^2\mmf_1 - t^2J\mmf_2\\
D_2^z(\omega) &=& J\mmf_2\omega^2 + J^3\mmf_1^2\mmf_2 - t^2J\mmf_1\\
T^0(\omega) &=& t\omega^2 + t^3 - tJ^2\mmf_1\mmf_2\\
T^z(\omega) &=& -it\omega J\left(\mmf_1+\mmf_2\right)
\end{eqnarray}
and the determinant is
\begin{eqnarray}
\text{det}\,A^\text{mf}(\omega) &=& \left[\omega^2+\frac{J^2}{2}(\mmf_1^2+\mmf_2^2)+t^2\right]^2 \notag\\
&&{}-\frac{J^4}{4}(\mmf_1^2-\mmf_2^2)^2 - t^2J^2(\mmf_1+\mmf_2)^2.\notag\\
\end{eqnarray}
Performing the trace in Eq.~\eqref{Eq:MagTrace} at $T=0$ then leads to the Lagrangian
\begin{eqnarray}
\lefteqn{\delta\mathcal{L}_\text{mag}(\Omega) =}\notag\\
&&-J^2\sum_{i=1,2}\Big[\left(D^{00}_i(\Omega)-D_i^{zz}(\Omega)\right)\mfl_i(\Omega)\cdot\mfl_i(-\Omega)\notag\\
&&\hspace{0.5cm}{}+2D^{zz}_i(\Omega)\tilde{m}_{i,z}(\Omega)\tilde{m}_{i,z}(-\Omega)\notag\\
&&\hspace{0.5cm}{}+i\left(D^{z0}_i(\Omega)-D^{0z}_i(\Omega)\right)\unitvec_z\cdot\left(\mfl_i(\Omega)\times\mfl_i(-\Omega)\right)\Big]\notag\\
&&{}-J^2\left(T^{00}(\Omega)-T^{zz}(\Omega)\right)\notag\\&&\times\left(\mfl_1(\Omega)\cdot\mfl_2(-\Omega)+\mfl_1(-\Omega)\cdot\mfl_2(\Omega)\right)\notag\\
&&{}-iJ^2\left(T^{z0}(\Omega)-T^{0z}(\Omega)\right)\notag\\&&\times\unitvec_z\cdot\left(\mfl_1(\Omega)\times\mfl_2(-\Omega)+\mfl_2(\Omega)\times\mfl_1(-\Omega)\right)\notag\\
&&{}-2J^2T^{zz}(\Omega)\left(\tilde{m}_{1z}(\Omega)\tilde{m}_{2z}(-\Omega)+\tilde{m}_{1z}(-\Omega)\tilde{m}_{2z}(\Omega)\right), \notag\\ \label{Eq:LongMagLagr}
\end{eqnarray}
with frequency $\Omega$, containing the integrals
\begin{equation}
D^{\alpha\beta}_i(\Omega) = \frac{1}{a^2}\int\!\frac{\text{d}\omega}{2\pi}\,\frac{D^\alpha_i(\omega)D^\beta_i(\omega-\Omega)}{\left[\text{det}\,A^\text{mf}(\omega)\right]\left[\text{det}\,A^\text{mf}(\omega-\Omega)\right]}
\label{Eq:DIntDefinition}
\end{equation}
and
\begin{equation}
T^{\alpha\beta}(\Omega) = \frac{1}{a^2}\int\!\frac{\text{d}\omega}{2\pi}\,\frac{T^\alpha(\omega)T^\beta(\omega-\Omega)}{\left[\text{det}\,A^\text{mf}(\omega)\right]\left[\text{det}\,A^\text{mf}(\omega-\Omega)\right]}.
\label{Eq:TIntDefinition}
\end{equation}
with $\alpha,\beta\in\{0,z\}$ and $i=1,2$. These integrals can be solved exactly by partial fraction decomposition, since the zeros of the denominator are known: $\text{det}\,A^\text{mf}(\omega)=0$ if $\omega^2=N^\pm$, with
\begin{eqnarray}
N^\pm &=& \pm J\sqrt{\frac{J^2}{4}(\mmf_1^2-\mmf_2^2)^2 + t^2(\mmf_1+\mmf_2)^2}\notag\\
&&{}-\frac12 J^2(\mmf_1^2+\mmf_2^2) - t^2,
\end{eqnarray}
where $N^-<0$ and $N^+\leq0$. Namely, $N^+=0$ if $t^2=J^2\mmf_1\mmf_2$, i.e., in terms of the dimensionless parameters, if $\tau=\mu$. This is where the discontinuity which is discussed in Sec.~\ref{Sec:NonTop} is located. %
In the integrals, we neglect terms of order $\Omega^2$ or higher in the long-wavelength limit, and obtain
\begin{eqnarray}
D^{00}_1(\Omega) &=& \frac{1}{4a^2\sqrt{-N^+}\left(N^+-N^-\right)^3}\notag\\
&&{}\times\Big[-(N^+)^3+5(N^+)^2N^-\notag\\
&&{}+2\left(J^2\mmf_2^2+t^2\right)\left((N^+)^2+3N^+N^-\right)\notag\\
&&{}+\left(J^2\mmf_2^2+t^2\right)^2\left(3N^++N^-\right)\Big]\notag\\
&&{}+(\text{same with }N^+\leftrightarrow N^-) + \mathcal{O}(\Omega^2)
\label{Eq:IntD00}
\end{eqnarray}
\begin{eqnarray}
D_i^{0z}(\Omega) &=& \frac{i\Omega}{16a^2N^+\sqrt{-N^+}(N^--N^+)^3}\notag\\
&&{}\times\Big[J\mmf_1(N^+)^2\left(2N^++9N^-\right)\notag\\
&&{}+J\mmf_1\left(J^2\mmf_2^2+t^2\right)N^+(N^+-5N^-)\notag\\
&&{}+J\mmf_2\left(J^2\mmf_2^2+t^2\right)\left(J^2\mmf_1\mmf_2-t^2\right)\notag\\
&&{}\times(10N^+-2N^-)\Big]\notag\\
&&{}+ (\text{same with }N^+\leftrightarrow N^-)+ \mathcal{O}(\Omega^3)
\label{Eq:IntD0z}
\end{eqnarray}
\begin{eqnarray}
D^{zz}_1(\Omega) &=& \frac{-J^2}{4a^2N^+\sqrt{-N^+}\left(N^+-N^-\right)^3}\notag\\
&&\times\Big[\mmf_1^2(N^+)^2\left(N^++3N^-\right)\notag\\
&&{}+ 2\mmf_1\mmf_2\left(J^2\mmf_1\mmf_2-t^2\right)N^+(3N^++N^-)\notag\\
&&{}+ \mmf_2^2\left(J^2\mmf_1\mmf_2-t^2\right)^2(5N^+-N^-)\Big]\notag\\
&&{}+(\text{same with }N^+\leftrightarrow N^-) + \mathcal{O}(\Omega^2)
\label{Eq:IntDzz}
\end{eqnarray}
\begin{eqnarray}
\label{Eq:T00}
T^{00}(\Omega) &=& \frac{-t^2\left(J^2\mmf_1\mmf_2-t^2-N^+\right)}{a^2\sqrt{-N^+}\left(N^+-N^-\right)^2}\notag\\
&&\times\left[1 + \frac{\left(5N^+-N^-\right)\left(J^2\mmf_1\mmf_2-t^2-N^+\right)}{4N^+\left(N^+-N^-\right)}\right]\notag\\
&& {}+(\text{same with }N^+\leftrightarrow N^-) + \mathcal{O}(\Omega^2)
\end{eqnarray}
\begin{eqnarray}
T^{0z}(\Omega) &=& \frac{i\Omega t^2(\mmf_1+\mmf_2)}{16a^2N^+\sqrt{-N^+}(N^--N^+)^3}\notag\\
&&{}\times\big[\left(t^2-J^2\mmf_1\mmf_2\right)\left(10N^++2N^-\right)\notag\\
&&{}-7(N^+)^2+N^+N^-\big]\notag\\
&&{}+(\text{same with }N^+\leftrightarrow N^-) + \mathcal{O}(\Omega^3)
\label{Eq:IntT0z}
\end{eqnarray}
\begin{eqnarray}
\label{Eq:Tzz}
T^{zz}(\Omega) &=&\frac{t^2J^2(\mmf_1+\mmf_2)^2\left(3N^++N^-\right)}{4a^2\sqrt{-N^+}\left(N^+-N^-\right)^3}\notag\\
&&{}+ (\text{same with }N^+\leftrightarrow N^-) + \mathcal{O}(\Omega^2)
\end{eqnarray}
Expressions for $D_2^{00}(\Omega)$, $D^{0z}(\Omega)$, and $D_2^{zz}(\Omega)$ can be obtained from Eqs.~\eqref{Eq:IntD00}, \eqref{Eq:IntD0z}, and \eqref{Eq:IntDzz}, respectively, by exchanging $\mmf_1\leftrightarrow\mmf_2$. It turns out that $D^{00}_1+D^{zz}_1=D^{00}_2+D^{zz}_2=-(T^{00}+T^{zz})$.
Furthermore, $D_i^{z0}(\Omega) = D_i^{0z}(-\Omega) = -D_i^{0z}(\Omega)$ and $T^{z0}(\Omega)=T^{0z}(-\Omega)=-T^{0z}(\Omega)$. These relations follow by substituting $\omega\rightarrow(\omega+\Omega)$ in Eqs.~\eqref{Eq:DIntDefinition} and \eqref{Eq:TIntDefinition} and the fact that only odd powers of $\Omega$ appear in Eqs.~\eqref{Eq:IntD0z} and \eqref{Eq:IntT0z}. %
For ease of notation, we write $D^{0z}_i(\Omega)=i\Omega \mathcal{D}^{0z}_i$ and $T^{0z}(\Omega)=i\Omega \mathcal{T}^{0z}$, where $\mathcal{D}^{0z}_i$ and $\mathcal{T}^{0z}_i$ are frequency-independent.

The effective magnetic surface Lagrangian that is evoked by the fermionic fluctuations, Eq.~\eqref{Eq:LongMagLagr}, is in real space and time given by
\begin{eqnarray}
\lefteqn{\delta\mathcal{L}_\text{mag}(\mathbf{r},t) =}\notag\\
&&-J^2\sum_{i=1,2}\Big[\left(D^{00}_i-D^{zz}_i\right)\mfl_i^2(\mathbf{r},t)+2D^{zz}_i\tilde{m}_{i,z}^2(\mathbf{r},t)\notag\\
&&{}-2\mathcal{D}^{0z}_i\unitvec_z\cdot\left(\mfl_i(\mathbf{r},t)\times\partial_t\mfl_i(\mathbf{r},t)\right)\Big]\notag\\
&&{}-2J^2(T^{00}-T^{zz})\mfl_1(\mathbf{r},t)\cdot\mfl_2(\mathbf{r},t)\notag\\
&&{}+2J^2\mathcal{T}^{0z}\unitvec_z\cdot\big[\mfl_1(\mathbf{r},t)\times\partial_t\mfl_2(\mathbf{r},t)\notag\\
&&{}+\mfl_2(\mathbf{r},t)\times\partial_t\mfl_1(\mathbf{r},t)\big]\notag\\
&&{}-4J^2T^{zz}\tilde{m}_{1z}(\mathbf{r},t)\tilde{m}_{2z}(\mathbf{r},t).
\label{Eq:dLmag} 
\end{eqnarray}
Equation \eqref{Eq:MagLagr} in section~\ref{Sec:NonTop} follows by writing the Lagrangian in terms of $\magn_i = \mmf_i\unitvec_z + \mfl_i$ again, where constant terms are discarded. The meaning of the different contributions is discussed in the main text.

In the special case of a pure AFM, where $\mmf_1=-\mmf_2$, a mathematical subtlety arises. Namely, the solution of the integrals $D_i^{\alpha\beta}(\Omega)$ and $T^{\alpha\beta}(\Omega)$ by partial fraction decomposition requires a different ansatz, because the zeros of the denominator are degenerate: $N^+=N^-=-J^2\mmf_1^2-t^2$. The integrals are notably easier as a consequence of multiple cancellations, and we find, again to leading order in $\Omega$ in the low-frequency regime,
\begin{equation}
D^{00}_{1,\text{AFM}}(\Omega) = D^{00}_{2,\text{AFM}}(\Omega) = -\frac{1}{4a^2\sqrt{J^2\mmf_1^2+t^2}} + \mathcal{O}(\Omega^2)
\end{equation}
\begin{equation}
D^{0z}_{1,\text{AFM}}(\Omega) = \!-D^{0z}_{2,\text{AFM}}(\Omega)\! = \frac{i\Omega J\mmf_1}{8a^2(J^2\mmf_1^2+t^2)^{3/2}} + \mathcal{O}(\Omega^3)
\end{equation}
\begin{equation}
D^{zz}_{1,\text{AFM}}(\Omega) = D^{zz}_{2,\text{AFM}}(\Omega) = \frac{J^2\mmf_1^2}{4a^2(J^2\mmf_1^2+t^2)^{3/2}} + \mathcal{O}(\Omega^2)
\end{equation}
\begin{equation}
T^{00}_\text{AFM}(\Omega) = \frac{t^2}{4a^2(J^2\mmf_1^2+t^2)^{3/2}} + \mathcal{O}(\Omega^2)
\end{equation}
\begin{equation}
T^{0z}_\text{AFM}(\Omega) = T^{zz}_\text{AFM}(\Omega) = 0.
\end{equation}
We have checked that these expressions are identical to the continuous limit $\mmf_2\rightarrow -\mmf_1$ of the integrals in the general case. Notably, no mixed Berry phase term is generated for the AFM. The fluctuation-induced Lagrangian takes the simplified form:
\begin{eqnarray}
\lefteqn{\delta\mathcal{L}_\text{mag}^\text{AFM} =}\notag\\ 
&&\frac{J^2\!\left[t^2\magn_1\!\cdot\magn_2
+ 2t^2\mmf_1(m_{1z}\!-m_{2z})
+ J^2\mmf_1^2(m_{1z}^2\!+m_{2z}^2)\right]}{-2a^2(J^2\mmf_1^2+t^2)^{3/2}}\notag\\
&&{}+\frac{J^3\mmf_1}{4a^2(J^2\mmf_1^2+t^2)}\unitvec_z\cdot(\magn_1\times\partial_t\magn_1 - \magn_2\times\partial_t\magn_2)
\end{eqnarray}

\section{Fluctuation-induced Landau theory}\label{App:LandauExpansion}
In this appendix, we present the Landau expansion of the energy in terms of the mean-field magnetizations at the interface. Here, we allow arbitrary directions of the magnetizations. Thus, the Landau theory is still valid if $\magn_1$ and $\magn_2$ are not aligned with each other or the $z$ axis at mean-field. For simplicity, we drop the overline-notation indicating mean-field values in this appendix.
\par The energy contains two contributions, namely (i) from the term $\det A$ in Eq.~\eqref{Eq:MagTrace}  originating with the quantum fluctuations of the sublattice fermions, and (ii) from a similar term $\det B$ generated by the quantum fluctuations of the Dirac fermions, where $B$ is defined such that Eq.~\eqref{Eq:EffDiracLagrangian1} can be written as $\mathcal{L}_\text{eff}=\Psi^\dagger B \Psi$. The energy density is then given by
\begin{equation}
\mathcal{E} = -\int\!\frac{d\omega}{2\pi}\int\!\frac{d^2k}{2\pi}(\ln\det A + \ln\det B)
,\end{equation}
where we use the cut-off value $\pi/a$ in divergent momentum integrals.
We did not include Landau terms for the bulk in Eq.~\eqref{Eq:L-Mag}, however, any bulk contributions would simply add up with the interface terms shown here. We obtain the following expansion to fourth order, where $\perp$ indicates the component orthogonal to the interface and $\parallel$ the in-plane component:
\begin{eqnarray}
\mathcal{E} &=& J^2\Bigg[\frac{-1}{4a^2|t|}(\magn_1-\magn_2)^2-t^2K_2(\magn_1+\magn_2)_\perp^2\notag\\
&&{}-\left(t^2K_2(1-v_F^2)+\frac{5h^4}{128\pi v_F^2|t|^3}\right)(\magn_1+\magn_2)^2_\shortparallel\Bigg]\notag\\
&&{}+J^4\big[c_1(m_1^4+m_2^4) + c_2m_1^2m_2^2+c_3(\magn_1\cdot\magn_2)^2\notag\\
&&{}+ c_4(m_1^2+m_2^2)\magn_1\cdot\magn_2+ c_5(m_1^2m_{1\shortparallel}^2+m_2^2m_{2\shortparallel}^2)\notag\\
&&{}+c_6(m_1^2m_{2\shortparallel}^2+m_2^2m_{1\shortparallel}^2)+c_7(\magn_1^2+\magn_2^2)(\magn_{1\shortparallel}\!\cdot\!\magn_{2\shortparallel})\notag\\
&&{}+c_8(m_{1\shortparallel}^2+m_{2\shortparallel}^2)\magn_1\cdot\magn_2+K_1(\magn_{1\shortparallel}+\magn_{2\shortparallel})^4\notag\\
&&{}+2c_8(\magn_{1\shortparallel}\cdot\magn_{2\shortparallel})(\magn_1\cdot\magn_2)\big].
\end{eqnarray}
The coefficients of the fourth-order terms are
\begin{eqnarray}
c_1 &=& \frac{1}{64a^2|t|^3} + K_1 +K_3 - K_4\\
c_2 &=& \frac{-7}{64a^2|t|^3} + K_1 + K_2 + K_3 - K_4\\
c_3 &=& \frac{5}{16a^2|t|^3} + 4K_1 - 4K_4\\
c_4 &=& \frac{-1}{16a^2|t|^3} +4K_1 +K_2 +2K_3 - 4K_4\\
c_5 &=& \frac{7h^4}{1024\pi v_F^2|t|^5} - 2K_1 -v_F^2(K_3-K_4)\\
c_6 &=& \frac{237h^4}{1024\pi v_F^2|t|^5} -2K_1 - v_F^2(K_2+K_3-K_4)\\
c_7 &=& \frac{47h^4}{512\pi v_F^2|t|^5} -4K_1 -v_F^2(K_2+2K_3-2K_4)\,\,
\end{eqnarray}
\begin{eqnarray}
c_8 &=& \frac{-63h^4}{512\pi v_F^2|t|^5} -4K_1 + 2v_F^2K_4
\end{eqnarray}
and we have used the constants
\begin{equation}
K_1 = \frac{6435h^8}{2^{15}\pi v_F^2|t|^9}
\end{equation}
\begin{widetext}
\begin{equation}
K_2 = h^4\frac{92\pi^2v_F^2 +108\pi v_Fa|t|+33a^2t^2}{48v_F|t|^5(2\pi v_F + a|t|)^3}+\frac{5h^4\log(1+\frac{2\pi v_F}{a|t|})}{64\pi v_F^2|t|^5}
\end{equation}
\begin{equation}
K_3 = h^4\frac{1408\pi^3v_F^3 + 2396\pi^2v_F^2a|t|+1392\pi v_Fa^2t^2+279a^3|t|^3}{384v_F|t|^5(2\pi v_F + a|t|)^4}
+\frac{35h^4\log(1+\frac{2\pi v_F}{a|t|})}{512\pi v_F^2|t|^5}
\end{equation}
\begin{equation}
K_4 = h^4\frac{9008\pi^4v_F^4 + 20000\pi^3v_F^3a|t|+16920\pi^2v_F^2a^2t^2+6500\pi v_Fa^3|t|^3+965a^4t^4}{1280v_F|t|^5(2\pi v_F + a|t|)^5}
+\frac{63h^4\log(1+\frac{2\pi v_F}{a|t|})}{1024\pi v_F^2|t|^5}
\end{equation}
\end{widetext}
It turns out that the second-order term is always negative, indicating a stable magnetic phase at the interface.
\par For the special cases of a FM, with $\magn_1=\magn_2=\mathbf{n}$, and an AFM, with $\magn_1=-\magn_2=\mathbf{n}$, the Landau theory can be simplified:
\begin{eqnarray}
\lefteqn{\mathcal{E}_\text{FM}}\notag\\
&=& -4J^2\left[t^2K_2n_\perp^2+\left(t^2K_2(1-v_F^2)+\frac{5h^4}{128\pi v_F^2|t|^3}\right)n_\shortparallel^2\right]\notag\\
&& + J^4\big[(16c_1+c_2+c_3+2c_4)n^4\notag\\
&&{}+ 2(c_5+c_6+c_7+2c_8)n_\shortparallel^2n^2+16K_1n_\shortparallel^4\big]
\end{eqnarray}
\begin{eqnarray}
\mathcal{E}_\text{AFM} &=& -\frac{J^2n^2}{a^2|t|}+ J^4\big[(16c_1+c_2+c_3-2c_4)n^4 \notag\\
&&{} + 2(c_5+c_6-c_7)n_\shortparallel^2n^2\notag\\
&&{}+16K_1n_\shortparallel^4\big]
\end{eqnarray}

\section{Landau-Lifshitz equation}\label{App:LLE}
Applying the Euler-Lagrange formalism on the total Lagrangian Eq.~\eqref{Eq:Lmag_tot} leads to the two equations of motion (with $i=1,2$ and $j=3-i$)
\begin{equation}
-\frac{\magn_i}{\magn_i^2}\times\partial_t\magn_i + b\unitvec_z\times\partial_t\magn_i
+ c\unitvec_z\times\partial_t\magn_j = \mathbf{d}_i. \label{Eq:eom}
\end{equation}
with the coefficients
\begin{equation}
b = 4J^2\mathcal{D}^{0z}_i-\frac{J_i^2}{4\pi v_F^2}
,\end{equation}
\begin{equation}
c = 4J^2\mathcal{T}^{0z}-\frac{J_1J_2}{4\pi v_F^2}
\end{equation}
and the effective field $\mathbf{d}_i = \mathbf{d}^i_\text{topol} + \mathbf{d}^i_\text{non-top}$ which consists of a part generated by the CS term,
\begin{eqnarray}
\mathbf{d}^i_\text{topol} &=&
\frac{eJ_i}{4\pi v_F}\mathbf{E}_\text{Cou}
+ \frac{eJ_i}{4\pi v_F}\mathbf{E}_\text{ext}
-\frac{\gamma\mmf_j}{4\pi v_F}(\nabla_\parallel\cdot\mathbf{M})\unitvec_z\notag\\
&&{}-\frac{\gamma J_i}{4\pi v_F}\nabla_\parallel(\mmf_1 m_{2z} + \mmf_2 m_{1z})
\end{eqnarray}
and the remainder containing various spin-stiffness and anisotropy terms besides the renormalized magnetic coupling of the sublattices
\begin{eqnarray}
\lefteqn{\mathbf{d}^i_\text{non-top} =}\notag\\
&&-\kappa\left(\nabla_\parallel\right)^2\magn_i
- \lambda\magn_j
- 4J^2 D_1^{zz}m_{1z}\unitvec_z\notag\\
&&{}- 2J^2\text{diag}(T^{00}-T^{zz},T^{00}-T^{zz},T^{00}+T^{zz})\cdot\magn_j\notag\\
&&{}+ 2J^2\left[(D^{00}_i+D^{zz}_i)\mmf_i+(T^{00}+T^{zz})\mmf_j\right]\unitvec_z\notag\\
&&{}+\frac{m_\Psi J_i}{\pi v_F^2}(J_1\mmf_1+J_2\mmf_2-M_z)\unitvec_z
-\frac{J_i}{12\pi m_\Psi v_F^2}\partial_t^2\mathbf{M}\notag\\
&&{}- \frac{J_i}{12\pi m_\Psi v_F}\partial_t\left[\gamma\nabla_\parallel(\mmf_1 m_{2z}+\mmf_2 m_{1z})-e\mathbf{E}_\text{ext}\right]\times\unitvec_z\notag\\
&&{}-\frac{J_i}{12\pi m_\Psi}\nabla_\parallel\left(\nabla_\parallel\cdot\mathbf{M}\right)\notag\\
&&{}- \frac{\gamma\mmf_j}{12\pi m_\Psi v_F}\left[\partial_t(\nabla_\parallel\times\mathbf{M})\cdot\unitvec_z\right]\unitvec_z\notag\\
&&{}+\frac{\gamma^2}{12\pi m_\Psi}\left(\nabla_\parallel\right)^2(\mmf_2^2 m_{1z} + \mmf_1^2 m_{2z})\unitvec_z\notag\\
&&{}+\frac{\gamma e}{12\pi m_\Psi}(\nabla_\parallel\cdot\mathbf{E}_\text{ext})\unitvec_z
- \frac{J_i}{12\pi m_\Psi}\left(\nabla_\parallel\right)^2M_z\unitvec_z\notag\\
\end{eqnarray}
with the short-hand notation $\mathbf{M}=J_1\magn_1+J_2\magn_2$. The second and third term in Eq.~\eqref{Eq:eom} are due to the fluctuation-induced Berry phases. Taking the cross product with $\mathbf{m}_i$ in Eq.~\eqref{Eq:eom}, using $\partial_t\magn_i^2=0$, one obtains
\begin{equation}
(1-bm_iz)\partial_t\magn_i - cm_{iz}\partial_t\magn_j + c(\magn_i\cdot\partial_t\magn_j)\unitvec_z = \magn_i\times\mathbf{d}_i.
\label{Eq:eom2}
\end{equation}
The equations of motion can now be rewritten in matrix form,
\begin{equation}
\Xi\cdot\begin{pmatrix}\partial_t\magn_1\\\partial_t\magn_2\end{pmatrix}
= \begin{pmatrix}\magn_1\times\mathbf{d}_1\\\magn_2\times\mathbf{d}_2\end{pmatrix},
\end{equation}
where the entries of the $(6\times6)$ matrix $\Xi$ follow from Eq.~\eqref{Eq:eom2}:
\begin{eqnarray}
\lefteqn{\Xi = \mathbb{1}_{(6\times6)}}\notag\\&&+
\begin{pmatrix}
-bm_{1z} & 0 & 0 & -cm_{1z} & 0      & 0 \\
0 & -bm_{1z} & 0 & 0      & -cm_{1z} & 0 \\
0 & 0 & -bm_{1z} & cm_{1x} & cm_{1y} & 0 \\
-cm_{2z} & 0      & 0 & -bm_{2z} & 0 & 0 \\
0      & -cm_{2z} & 0 & 0 & -bm_{2z} & 0 \\
cm_{1x} & cm_{2y} & 0 & 0 & 0 & -bm_{2z}
\end{pmatrix}\notag\\
\end{eqnarray}

%

\end{document}